\documentclass[runningheads]{llncs}
\usepackage[T1]{fontenc}
\usepackage{graphicx}
\usepackage[utf8]{inputenc} 
\usepackage{amsmath} 
\usepackage{amssymb}
\usepackage{graphicx} 
\usepackage{hyperref} 
\usepackage{listings}
\usepackage{xcolor}
\usepackage{subcaption}

\definecolor{codegreen}{rgb}{0,0.6,0}
\definecolor{codegray}{rgb}{0.5,0.5,0.5}
\definecolor{codepurple}{rgb}{0.58,0,0.82}
\definecolor{backcolour}{rgb}{0.95,0.95,0.92}

\lstdefinestyle{mystyle}{
    backgroundcolor=\color{backcolour},   
    commentstyle=\color{codegreen},
    keywordstyle=\color{magenta},
    numberstyle=\tiny\color{codegray},
    stringstyle=\color{codepurple},
    basicstyle=\ttfamily\scriptsize,  
    breakatwhitespace=false,         
    breaklines=true,                 
    captionpos=b,                    
    keepspaces=true,                 
    numbersep=5pt,                  
    showspaces=false,                
    showstringspaces=false,
    showtabs=false,                  
    tabsize=2
}

\begin{document}
\title{\textsf{OSVAuto}: automatic proofs about functional specifications in OS verification}
%
%
\author{Yulun Wu\inst{1} \and
Bican Xia\inst{1} \and
Jiale Xu\inst{2,3} \and
Bohua Zhan\inst{4} \and
Tianqi Zhao\inst{5}}
%
%
\institute{School of Mathematical Sciences, Peking University, China \and
Institute of Software, Chinese Academy of Sciences, China \and
University of Chinese Academy of Sciences, China \and
Huawei Technologies Co., Ltd., China \and
Zhongguancun Laboratory, Beijing, China}
\maketitle              
\begin{abstract}
We present \textsf{OSVAuto} for automatic proofs about functional specifications that commonly arise when verifying operating system kernels. The algorithm behind \textsf{OSVAuto} is designed to support natively those data types that commonly occur in OS verification, including sequences, maps, structures and enumerations. Propositions about these data are encoded into a form that is suitable for SMT solving. For quantifier instantiation, we propose an extension of recent work for automatic proofs about sequences. We evaluate the algorithm on proof obligations adapted from existing verification of the $\mu$C-OS/II kernel in Coq, demonstrating that a large number of proof obligations can be solved automatically, significantly reducing the proof effort on the functional side.

\end{abstract}
\section{Introduction}

Verification of operating system kernels is one of the major application areas of interactive and automatic theorem proving. Major efforts using interactive theorem provers include the verification of the seL4 microkernel~\cite{DBLP:journals/tocs/KleinAEMSKH14}, CertiKOS~\cite{DBLP:conf/osdi/GuSCWKSC16}, and $\mu$C/OS-II~\cite{DBLP:conf/cav/XuFFZZL16}. Recent efforts such as~\cite{DBLP:conf/sosp/NelsonSZJBTW17,DBLP:conf/sosp/NelsonBGBTW19} place a greater focus on automation, making more extensive use of SMT solvers~\cite{DBLP:conf/tacas/MouraB08}. While there is a great variety of techniques and program logics involved in the different projects, there is also similarity in the general approach. Usually, specification of the kernel is described using a functional language at various levels of abstraction. They include abstract behavior of each API function, invariants on the internal data structures, and refinement relations between different abstraction levels. The verification task is then roughly divided into two parts:

\begin{enumerate}
\item \textbf{Correctness of functional specification:} including preservation of invariants, specification at different abstraction levels observe the refinement relation, and that the specification at the most abstract level satisfy important system properties (such as correctness of task scheduling).

\item \textbf{Refinement between functional specification and implementation:} This involves reasoning about semantics of low-level programming languages, in particular issues of memory management and aliasing, and is often done using some version of (concurrent) separation logic.
\end{enumerate}

In this paper, we focus on providing greater level of automation to the first part. In real verification projects, the invariant of the OS kernel can be highly complex, and hence the proof of invariant preservation and refinement relations take up a large proportion of the proof effort~\cite{DBLP:journals/tocs/KleinAEMSKH14}.

A key observation of this paper is that in the domain of OS verification, the nature of proof obligations that arise is actually quite restricted. Apart from lists (which can be encoded in SMT using sequence theory), most of the data structures that arise are non-recursive. Unlike proving theorems in mathematics or about complex algorithms, there is little need to make use of a large body of existing lemmas. While the invariants frequently involve quantifiers, the quantification is usually over indices of arrays or keys of mappings, making quantifier instantiation techniques for sequences highly applicable.

In particular, recent work~\cite{DBLP:journals/jar/WangA23} proposed proof automation techniques for sequences, where quantifier instantiation is determined by analyzing the appearances of bound variables within array indices. Within the restricted language of~\cite{DBLP:journals/jar/WangA23}, it is able to produce a finite number of instantiations when successful, resulting in a proof goal that is equisatisfiable with the original (the method is both \emph{sound} and \emph{complete}). It also provides quick checks for failure (by detecting nonzero cycles in the classification graph). While the algorithm in~\cite{DBLP:journals/jar/WangA23} has good theoretical properties, it is limited on the practical side by the following restrictions:

\begin{itemize}
\item Only one-dimensional sequences are supported, but nested sequences (and maps) frequently arise when modeling OS data structures.
\item Structures and enumerations are not supported. They also appear frequently in modeling, often nested with sequences and maps.
\item Many common properties on sequences cannot be handled, for example uniqueness of elements in a sequence.
\end{itemize}

In this paper, we propose an extension of the algorithm in~\cite{DBLP:journals/jar/WangA23} that relaxes the above restrictions. In particular, we include support for nested sequences, (finite) mappings, structures and enumerations. To deal with difficulties mainly arising from nested sequences, we propose keeping track of \emph{conditions} on instantiations when propagating along the graph, as well as new propagation rules for sequence and map nodes having the same form.

Relaxing the above restrictions make the theoretical issues significantly more difficult, so we do not aim for completeness of our algorithm, but rather impose heuristic cutoffs on the instantiation when it diverges. Some of the issues with completeness are discussed in Section~\ref{sec:discussion}.

We evaluate the algorithm on functional specification and proof obligations adapted from the existing verification of the $\mu$C/OS-II kernel in Coq~\cite{DBLP:conf/cav/XuFFZZL16}. We observe that while the instantiation algorithm is not theoretically complete, it is able to find all the necessary instantiations automatically, on a large number of examples. This results in potentially significant reduction in proof effort on the functional side.

In summary, the main contributions of this paper is as follows:
\begin{itemize}
\item We describe an extended language for writing functional specifications that is well-suited for OS verification, including native support for sequences, mappings, structures and enumerations.
\item We extend the quantifier instantiation algorithm for sequences to the new language, with novel techniques including keeping track of conditions during propagation, and propagating along nodes having the same form.
\item We evaluate the algorithm on the existing $\mu$C/OS-II verification in Coq, demonstrating that it can achieve complete automation on many proof obligations.
\end{itemize}

\section{Preliminaries}

In this section, we review existing works in proof automation for sequence theory, and the existing verification of the $\mu$C/OS-II kernel.

\subsection{Automation of sequence theory}
\label{sec:sequence-preliminary}

Our work extends the algorithm in~\cite{DBLP:journals/jar/WangA23} for proof automation in sequence theory. We sketch a general idea of the algorithm here.

The main problem addressed is how to compute a finite set of instantiations for the quantified facts in the proof obligation, such that the instantiated version of the goal is equisatisfiable to the original (the instantiations are both \emph{sound} and \emph{complete}). The algorithm proceeds by constructing a graph where each node represents either a bound variable of some quantified fact, or a sequence variable. An appearance of a term $a[i+c]$ in the body of quantification over $i$ induces an edge from $i$ to $a$ with weight $c$ (here $c$ does not contain any bound variables. There are other rules for inducing edges which we omit here). The algorithm is successful if there are no cycles with nonzero total weight in the resulting graph. Some initial instantiations are induced by concrete indices of the form $a[k]$, inequality conditions, etc. Other instantiations are obtained by traversing the graph and adding edge weights to existing instantiations.

A number of operations on sequences are supported natively, including \textsf{append}, \textsf{slice}, \textsf{repeat} and \textsf{map}. Appearances of such operations are reduced by simplifying \textsf{length} and \textsf{index} terms on the resulting sequences, with case splitting performed if necessary. For example, an appearance of $\textsf{append}(a,b)[i]$ is reduced to $a[i]$ under $0\le i<|a|$ and $b[i-|a|]$ under condition $|a|\le i<|a|+|b|$ \footnote{there is also a case for default value where index is out of bounds, which we omit in this paper throughout.}.

We borrow the main example from~\cite{DBLP:journals/jar/WangA23} to illustrate the algorithm. A variant of this example will be used in Section~\ref{sec:examples} to illustrate one of our extensions. Given sequences $a,b,c,d$ and predicate $P$, assumptions $c=\textsf{append}(a,b)$, $d=\textsf{append}(b,a)$, we wish to show
\[ (\forall i.\, 0\le i<|c| \rightarrow P(c[i])) \Longrightarrow (\forall k.\, 0\le k<|d| \rightarrow P(d[k])) \]
The first step of the algorithm makes substitutions $c=\textsf{append}(a,b)$ and $d=\textsf{append}(b,a)$, followed by reducing appearances of \textsf{append}. Eventually this results in proof goals such as the following:
\begin{align*}
0\le k<|b| &\Longrightarrow (\forall i.\, 0\le i<|a| \rightarrow P(a[i])) \\
&\Longrightarrow (\forall j.\, |a|\le j<|a| + |b| \rightarrow P(b[j-|a|])) \Longrightarrow P(b[k])
\end{align*}
This proof goal gives rise to a graph with nodes $i, j, a$ and $b$, with edge $i\to a$ of weight 0 (from $a[i]$) and $j\to b$ of weight $-|a|$ (from $b[j-|a|]$). Initial instantiation $k$ for node $b$ comes from the term $b[k]$ in the goal. This induces instantiation $k+|a|$ for bound variable $j$. With this instantiation, the proof goal becomes:
\begin{align*}
0\le k<|b| &\Longrightarrow (|a|\le k+|a|<|a|+|b| \to P(b[k])) \Longrightarrow P(b[k])
\end{align*}
which is clearly provable.

\subsection{Verification of the $\mu$C/OS-II kernel}
\label{sec:ucos-preliminary}

In this section, we give an overview of the existing verification of the $\mu$C/OS-II operating system in Coq~\cite{DBLP:conf/cav/XuFFZZL16}. The purpose is to provide a background for the current work, showing the kind of specifications and proof goals that commonly arise in OS verification, and motivate the language and algorithm designs in later sections.

The $\mu$C/OS-II is a small OS kernel, whose core modules include task management, event management, timers, and so on. Existing work~\cite{DBLP:conf/cav/XuFFZZL16} verified in Coq most of the key functionalities in the kernel. The verification makes use of concurrent separation logic with refinement (CSL-R) that handles preemptive interrupts. Functional specification is defined for each API function at two levels of abstraction (called high-level and low-level). Invariants are stated for data structures at each level, and refinement relations are stated between the two levels. They will be collectively called \emph{invariants} since they are preserved by the combination of high-level and low-level functional specification of each API function.

We now give some examples of data structures and invariants. The definitions largely follow the existing Coq verification, but we will show their restatement in the language of \textsf{OSVAuto} for ease of reading.

\textsf{Address}es are modeled using unsigned 32-bit integers. Each address value (\textsf{addrval}) is either null (\textsf{Vull}) or a pointer value (\textsf{Vptr(addr)}). The low-level task control block (\textsf{TCB}) contains information about a task, including its priority, status (represented as a bitmap), event it is waiting for, etc, represented in a way that is close to the C implementation. By comparison, the high-level TCB (\textsf{AbsTCB}) contains the same information but in a more abstract way, representing status as an enumerated value. These definitions are shown in Figure~\ref{fig:example-tcb}.

\begin{figure}[h!]
\begin{subfigure}{0.42\textwidth}
\begin{lstlisting}[language=C]
// Addresses
typedef address = int32u;

enum addrval =
  Vnull | Vptr(address addr)

// Low-level TCB
struct TCB {
  addrval OSTCBEventPtr;
  addrval OSTCBMsg;
  int16u OSTCBDly;
  int16u OSTCBStat;
  int8u OSTCBPrio;
  int8u OSTCBX;
  int8u OSTCBY;
  int8u OSTCBBitX;
  int8u OSTCBBitY;
}
\end{lstlisting}
\end{subfigure}
\begin{subfigure}{0.58\textwidth}
\begin{lstlisting}[language=C]
// High-level task status
enum tcbstats =
  os_stat_sem (address ev)
| os_stat_q (address ev)
| os_stat_time
| os_stat_mbox (address ev)
| os_stat_mutexsem (address ev)

enum taskstatus =
  rdy | wait(tcbstats stat, int16u time)

// High-level TCB
struct AbsTCB {
  int8u prio;
  taskstatus stat;
  addrval msg;
  bool sus;
}
\end{lstlisting}
\end{subfigure}
\caption{Examples of high-level and low-level task control blocks.}
\label{fig:example-tcb}
\end{figure}

Next, we consider refinement relations between the low-level and high-level TCBs. They express, for example, the correspondence between representation of task status, with one case for each possible status value. The case where the task is suspended and at the same time waiting for semaphore is shown in Figure~\ref{fig:example-refinement-tcb}.

\begin{figure}[h!]
\begin{lstlisting}[language=C]
predicate RHL_WaitS_Suspend_P(TCB tcb, Seq<int8u> rtbl, AbsTCB abstcb) {
  switch (abstcb) {
    case AbsTCB{{prio:prio, stat: wait(os_stat_sem(eid),dly), sus: true}}:
      tcb.OSTCBPrio == prio && prio_not_in_tbl(prio, rtbl) &&
      tcb.OSTCBEventPtr == Vptr(eid) &&
      tcb.OSTCBStat == OS_STAT_SEM | OS_STAT_SUSPEND;
    default: true;
  }
}
\end{lstlisting}
\caption{Part of the refinement relation for task control blocks (task is suspended while also waiting for a semaphore).}
\label{fig:example-refinement-tcb}
\end{figure}

The low-level and high-level global state (with type \textsf{Global} and \textsf{AbsGlobal} respectively) each contains a table mapping priorities to TCBs. The high-level state contains further information, for example a priority map for efficient retrieval of the ready task with highest priority. We omit their definitions for reason of space. The refinement relations on TCBs extend to refinement relations on the global state \texttt{RLH\_TCB\_P}.


Finally, we give an example of proof obligation from verifying the API function for suspending a task (Figure~\ref{fig:example-proof-obligation}).



\begin{figure}[h!]
\begin{lstlisting}[language=C]
function OSTaskSuspendAbs(int8u prio, AbsGlobal absGlobal) -> AbsGlobal {
  if (!indom(prio, absGlobal.tcbMap)) { absGlobal }
  else { absGlobal{|tcbMap[prio].sus := true|} }
}

query RLH_TCB_P_suspend3 {
  ... variable declarations ...
  assumes global2 == OSTaskSuspend(prio, global);
  assumes absGlobal2 == OSTaskSuspendAbs(prio, absGlobal);
  assumes prio >= 0 && prio < 64;
  assumes H1: RLH_TCB_P(global, absGlobal);
  assumes H2: RL_TCB_P(global);
  assumes H3: RL_Tbl_Grp_P(global);
  shows RLH_TCB_P(global2, absGlobal2)
}
\end{lstlisting}
\caption{An example of proof obligation from verifying API function for suspending a task. The predicates \texttt{RL\_TCB\_P} and \texttt{RL\_Tbl\_Grp\_P} are invariants on the low-level global state. \texttt{OSTaskSuspend} and \texttt{OSTaskSuspendAbs} are functional specifications for suspend on the low-level and high-level respectively. Some definitions are omitted for reason of space.}
\label{fig:example-proof-obligation}
\end{figure}

As can be guessed from the above code samples, the full invariant and refinement relation is quite lengthy, and proof of invariant preservation can be very tedious. However, the proof obligations are not very deep in the sense of requiring substantial insights, rather they involve a lot of case analysis and mostly routine reasoning about sequences and maps. Such proof obligations should be well-suited for proof by SMT solvers. However, applying SMT to the original statements in Coq (e.g. using tools such as SMTCoq~\cite{DBLP:conf/cav/EkiciMTKKRB17}) faces the following difficulties:

\begin{itemize}
\item Tools such as SMTCoq do not encode lists and partial maps directly into corresponding SMT theories, in particular not making use of recent advances about proof automation in sequence theory.
\item Likewise, encoding for records and inductive definitions in Coq, which aim for generality, do not take advantage of the special nature of the problem to give a encoding that is efficient for SMT solving.
\item There are many extraneous quantifiers in the original Coq statements, for example to express matching, that are conventional in an interactive theorem prover but unsuitable for SMT solving.
\end{itemize}

The above difficulties make it simply impractical to directly use existing connection between Coq and SMT solvers (similar issues arise for hammer-like tools in other proof assistants). Hence, we design \textsf{OSVAuto} with the following aims: native representation and encoding for structures, enumerations, sequences and maps, and making full use of proof automation (in particular quantifier instantiation) techniques for the resulting theories.

\section{Language}
\label{sec:language}

In this section, we describe the language for expressing functional specification and proof obligations in \textsf{OSVAuto}. Frequent references will be made back to Section~\ref{sec:ucos-preliminary} which give examples of using various constructs of the language.

\subsection{Types}
\label{sec:language-types}

The primitive types include \textsf{bool}, \textsf{int}, and signed/unsigned bit-vector types (\textsf{int8}, \textsf{int8u}, \textsf{int16}, \textsf{int16u}, etc). Structures are defined by a list of (named and typed) fields. Enumerations are defined by a list of alternatives, with each alternative given by a list of fields. If a name appears in more than one branch of an enumeration, it must have the same type in all branches. Enumerations correspond to types defined using \textsf{inductive} in Coq and \textsf{datatype} in Isabelle, except we do not allow recursion in the definitions. This is an important restriction which enables more efficient SMT encoding, as described in Section~\ref{sec:smt-encoding}.

Sequence types have the form $\textsf{Seq<}T\textsf{>}$, where $T$ is the type of entries of the sequence. Map types have the form $\textsf{Map<}K, V\textsf{>}$, where $K$ and $V$ are key and value types, respectively. It is interpreted as a partial function from $K$ to $V$ with finite domain. We only support keys with integer and bit-vector types (there is no constraint on types of values).

Type definitions are supported (e.g. defining \textsf{address} to be \textsf{int32u} in Figure~\ref{fig:example-tcb}). They are expanded automatically at parsing time.

\subsection{Terms}
\label{sec:language-terms}

We support the following term constructors: variables, constants, operators (logical and arithmetic), function application (including defined functions and predicates, constructors of enumeration, conversion functions between integer and bit-vector types, and native operations on sequences and maps).

For structures, we provide term constructor for structure literals (e.g. \textsf{Point\{x:} 3, \textsf{y}: 4\textsf{\}}) and structure update (e.g. $p\textsf{\{x} := 5\textsf{\}}$, where $p$ has type \textsf{Point}). Dot-notation is used for field access in structures (e.g. $p\textsf{.x}$ and $p\textsf{.y}$).

For enumerations, concrete values are built using function application. Dot-notation can be used to access fields of a enumeration. For example, for the enumeration \textsf{addrval} in Figure~\ref{fig:example-tcb}, $\textsf{Vptr}(10)\textsf{.addr}$ reduces to 10, while \textsf{Vnull.addr} is not meaningful (reduces to the default value of \textsf{int32u}). The special notation \textsf{.id} is used to access the branch number of an enumeration value. For example, $\textsf{Vptr}(10)\textsf{.id}$ reduces to 1 while \textsf{Vnull.id} reduces to 0.

More commonly the \textsf{switch} syntax is used to work with enumeration values. For example, the term
\[ \textsf{switch}\ (a)\ \{ \textsf{case Vnull} \Rightarrow 0; \textsf{case Vptr}(n) \Rightarrow n + 1; \} \]
reduces to 0 if $a$ is \textsf{Vnull}, and $n+1$ if $a$ has form $\textsf{Vptr}(n)$. In fact, \textsf{OSVAuto} supports a more general form \textsf{switch}, where the pattern can contain (partial) structure literals and constants. An example of its usage is shown in Figure~\ref{fig:example-refinement-tcb}, where the pattern matches \textsf{AbsTCB} values where the \textsf{sus} field is \textsf{true}, and the \textsf{stat} field has form $\textsf{wait}(\textsf{os\_stat\_mutexsem}(\textit{eid}), \textit{dly})$. The variables \textit{prio}, \textit{eid} and \textit{dly} are bound and can appear in the body of the case.

A \textsf{let} term has syntax $\textsf{let}\ v = \textit{rhs}\ \textsf{in}\ \textit{body}\ \textsf{end}$, with standard meaning. Quantifiers \textsf{forall} and \textsf{exists} are supported. As syntax sugar, one can optionally specify bounds on the quantifier (as integer ranges or keys of a mapping).

Sequence types support two \emph{basic} operations: \textsf{index} for index access (with syntax $a[i]$ used for $\textsf{index}(i, a)$) and \textsf{length} for length of a sequence ($|a|$ is used for $\textsf{length}(a)$). The \emph{non-basic} operations \textsf{append}, \textsf{cons}, \textsf{update}, \textsf{slice}, \textsf{repeat}, and \textsf{remove} can be used in the specification, but will be eliminated during normalization in favor of basic ones. Map types have two basic operations \textsf{indom} and \textsf{get}, and two non-basic operations \textsf{empty} and \textsf{update} for creating new maps.

When writing functional specification, it is often useful to express updates deep inside a structure, for example certain index of a certain field (and so on). We support a general syntax of structure update for this purpose. A canonical example is given in Figure~\ref{fig:example-proof-obligation}, where \textit{absGlobal}$\{$\textsf{tcbMap}[\textit{prio}]\textsf{.sus} := \textsf{true}$\}$ means updating the \textsf{sus} field in the \textit{prio}'th element of field \textsf{tcbMap} in \textit{absGlobal} to \textsf{true}.

\subsection{Function definitions and proof obligations}

In \textsf{OSVAuto}, \emph{function} is defined using syntax starting with keyword \textsf{function}. Types of each parameter and return value must be specified. \emph{Predicate} is just a special case of function where the return value is \textsf{bool}. Recursive functions are not allowed. This restriction is made to enable unfolding all function definitions during normalization (Section~\ref{sec:normalization}).

Each proof obligation is specified using syntax starting with \textsf{query}. They may contain type variables, variables, assumptions, and a goal. Examples of function definition and proof obligation are shown in Figure~\ref{fig:example-proof-obligation}.

\subsection{Generalized atomic terms}
\label{sec:generalized-atomic}

We now introduce a concept that will be used frequently when describing proof automation later on.

\begin{definition}
The set of \emph{generalized atomic terms} is constructed inductively from the following rules:
\begin{itemize}
\item A variable is a generalized atomic term.
\item If $s$ is generalized atomic with structure or enumeration type, then field accesses (including \textsf{.id}) on $s$ are generalized atomic.
\item If $a$ is generalized atomic with sequence or map type, then the basic operations $a[i]$, $|a|$, $\textsf{indom}(k, a)$ and $\textsf{get}(k, a)$ are generalized atomic. Note there are no constraints on indices $i$ or keys $k$.
\end{itemize}
\end{definition}

Each generalized atomic term can be decomposed into a pair $(\textit{name}, \textit{idx})$, where \textit{name} is a dot-separated string, each part being either name of a variable in the current goal, field name (including \textsf{id}), or one of the four basic operations \textsf{index}, \textsf{length}, \textsf{get} and \textsf{indom}. The second part \textit{idx} record the list of indices/keys for \textsf{index}, \textsf{get} and \textsf{indom}. We illustrate with some examples:

\begin{example}
\label{ex:atomic}
Suppose the following structures and enumerations are declared:
\begin{align*}
&\textsf{struct}\ \textsf{Point}\ \{ \textsf{int x}; \textsf{int y}\} \\
&\textsf{enum}\ \textsf{Shape} = \textsf{polygon}(\textsf{Seq<Point> pts}) \ |\ \textsf{single}(\textsf{Point pt})
\end{align*}
Given variables $s$ of type \textsf{Shape} and $m$ of type $\textsf{Map<int, Shape>}$, some examples of generalized atomic terms and their decompositions as (\textit{name}, \textit{idx}) pairs are:
\begin{itemize}
\item $s.\textsf{pt}.\textsf{y}$, decomposed as $(\textsf{s.pt.y}, [])$.
\item $s.\textsf{pts}[i].\textsf{x}$, decomposed as $(\textsf{s.pts.index.x}, [i])$.
\item $|s.\textsf{pts}|$, decomposed as $(\textsf{s.pts.length}, [])$.
\item $m[k].\textsf{id}$, decomposed as $(\textsf{m.get.id}, [k])$.
\item $m[k].\textsf{pts}[i].\textsf{y}$, decomposed as $(\textsf{m.get.pts.index.y}, [k, i])$.
\end{itemize}
\end{example}

Note both the length of \textit{idx} and the type of each entry in \textit{idx} depend on \textit{name} only (by looking for appearances of \textsf{index}, \textsf{get}, \textsf{indom} in \textit{name} from left to right). The type of the generalized atomic term can also be determined from \textit{name} and types of variables in the goal.

\section{Algorithm}
\label{sec:algorithm}

In this section, we present the proof automation algorithm in \textsf{OSVAuto}. On a high-level, the algorithm consists of the following four steps:

\begin{enumerate}
\item \textit{Normalize:} the goal is reduced by applying normalization rules, until it is made up of generalized atomic terms and a limited set of syntax elements.
\item \textit{Instantiation:} facts that are universally quantified are instantiated, eventually reducing the goal into quantifier-free form.
\item \textit{Encoding:} the quantifier-free proof goal is encoded into SMT formula involving only uninterpreted functions and arithmetic theories.
\item \textit{Solve:} finally, the SMT formula is checked for satisfiability using the backend SMT solver (we use \textsf{z3} for our experiments).
\end{enumerate}

At any stage, we apply the first step in the above list that is applicable. That is, after any instantiation step, we check whether the result can be further normalized, and we proceed to encoding only after no further instantiation can be performed.

Among the above steps, 1 and 3 preserve the satisfiability of the proof goal. In step 2, the choice of instantiation is \emph{heuristic}, in the sense that it is not guaranteed that all necessary instantiations for the proof are found. Hence, the entire algorithm is \emph{sound} but not \emph{complete}. Nevertheless, we justify the method of instantiation based on intuitive principles, following the ideas in~\cite{DBLP:journals/jar/WangA23}, and we show in experiments that they are sufficient for many proof obligations that arise in practice in OS verification.

\subsection{Normalization}
\label{sec:normalization}

The first step reduces the goal using a number of normalization rules. Most of these are standard, so we will not explain them further.

\begin{enumerate}
\item All function (and predicate) definitions are expanded.
\item General structure updates (discussed at end of Section~\ref{sec:language-terms}) are expanded in terms of basic structure, sequence and map updates.
\item (Generalized) \textsf{switch} statements are expanded into \textsf{if-then-else} statements.
\item \textsf{let}-terms on the top-level are removed by creating a new variable $v$ and adding the corresponding equality. Otherwise they are expanded.
\item Equality between terms of structure, enumeration, sequence and map types are expanded according to their definitions.
\item Bounded quantifiers are converted into unbounded ones by moving the constraint into the body.
\item Quantifiers are moved to the outermost level (prenex form). New variables are created from existence facts on the outermost level (skolemization).
\item The non-basic sequence and map operations are reduced according to their semantics. However, we do not split assumptions and proof goals according to conditions, but the represent different cases as \textsf{if-then-else} terms. This keeps the number of assumptions and proof goals small. Part of the advantages of splitting described in~\cite{DBLP:journals/jar/WangA23}, that they remove non-zero cycles in some cases, can be achieved with more generality by keeping track of conditions during propagation, as discussed in Section~\ref{sec:instantiation}.
\end{enumerate}

A further optimization, making use of \emph{defining equations}, is adopted from~\cite[Section 3.1]{DBLP:journals/jar/WangA23}. We say a fact is a defining equation if it is an equality whose left side is a generalized atomic term with no indices. The above normalization is applied only to the indices and right side of defining equations, and the equations themselves are used for rewriting in the rest of the proof goal.

The following result is shown by a routine analysis of the above normalization procedure.

\begin{theorem}
\label{thm:normalization}
The above normalization procedure results in a proof goal consisting only of generalized atomic terms, constants, logical and arithmetic operators (including \textsf{if-then-else}), and quantifiers. Equalities are applied only to terms of primitive type.
\end{theorem}

In particular, all non-basic operations on sequences and maps are removed, all structure literal, structure value, and enumeration constructors are removed, and basic operations for sequence/maps and field accesses are only applied to generalized atomic terms (and hence form part of a generalized atomic term).

Part 8 of normalization follow the method in~\cite{DBLP:journals/jar/WangA23}, and we refer to that paper for details and examples. We illustrate step 5 using declarations from Example~\ref{ex:atomic}.

\begin{example}
\label{ex:normalize}
Consider the equality $s=t$ where $s,t$ have type \textsf{Shape}. This is normalized into the following:
\begin{align*}
&s.\textsf{id} = t.\textsf{id}\,\wedge \\
&(s.\textsf{id} = 0 \longrightarrow |s.\textsf{pts}| = |t.\textsf{pts}|\,\wedge \\
& \quad\quad\forall i.\ 0\le i<|s.\textsf{pts}| \longrightarrow s.\textsf{pts}[i].\textsf{x} = t.\textsf{pts}[i].\textsf{x} \wedge s.\textsf{pts}[i].\textsf{y} = t.\textsf{pts}[i].\textsf{y}) \,\wedge \\
&(s.\textsf{id} = 1 \longrightarrow s.\textsf{pt}.\textsf{x} = t.\textsf{pt}.\textsf{x} \wedge s.\textsf{pt}.\textsf{y} = t.\textsf{pt}.\textsf{y})
\end{align*}
\end{example}

\subsection{Quantifier instantiation}
\label{sec:instantiation}

After normalization, the main construct that still need to be removed before SMT solving is universal quantifiers in facts. The quantifier instantiation procedure largely follows the ideas of~\cite{DBLP:journals/jar/WangA23}, by building and analyzing a \emph{classification graph} (called \emph{graph} for short). We extend the method to handle nested sequences and maps (issues about structures and enumerations have already been addressed by normalization). The main modifications are summarized as follows:
\begin{itemize}
\item In addition to variables, nodes in the graph can be created for any generalized atomic terms of sequence or map type.
\item The quantifiers are instantiated one-at-a-time, starting from the outermost.
\item We keep track of \emph{conditions} for each edge in the graph and each instantiation. Propagation of instantiations stop when the accumulated conditions become unsatisfiable.
\item We add propagation of instantiations along \emph{pattern edges}, and between any two sequence or map nodes with the same name.
\end{itemize}

With the above extensions, we choose the formulate the algorithm in terms of \emph{propagation} of instantiations along edges of the graph (this is equivalent to the formulation in terms of index sets in the simplified setting of~\cite{DBLP:journals/jar/WangA23}). We also need to abandon the quick detection of failure using loops with nonzero weights, instead add heuristic cutoffs in the number and generation of instantiations.

We now describe the instantiation procedure, starting from the basic definitions. The classification graph is initially empty, and at any stage, the following nodes are added to the graph:
\begin{itemize}
\item Node $i$ is added for each universally quantified fact, whose outermost bound variable is $i$.
\item Node $s$ is added for each generalized atomic term $s$ that is \emph{free} in the current proof goal (and for which no node is already present).
\end{itemize}

Both edges in the graph and instantiations are associated with a list of \emph{conditions}. First, we define the concept of conditions associated to a subterm $t'$ of a term $t$. Given a fixed term $t$,
\begin{itemize}
\item Conditions associated to $t$ itself is the empty list.
\item If the subterm $\textsf{if}\ (c)\ \{ t_1 \}\ \textsf{else}\ \{ t_2 \}$ has conditions $\textit{cs}$, then $t_1$ has conditions $\textit{c}::\textit{cs}$, and $t_2$ has conditions $\neg\textit{c}::\textit{cs}$.
\item If the term $c\to t$ has conditions $\textit{cs}$, then $t$ has conditions $c::\textit{cs}$.
\end{itemize}
In all other cases, the subterm inherit conditions from its parent term. Intuitively, the conditions specify the constraints under which the subterm need to be considered.

Now we are ready to describe how edges in the graph are formed. Edges come in two types: \emph{weighted edges} and \emph{pattern edges}.

\begin{itemize}
\item An edge with weight $c$ is added from bound variable $i$ to sequence $a$ if $a[i+c]$ appears in the body of quantifier over $i$. The conditions for the edge are that associated to $a[i+c]$ relative to the body of the quantified fact.
\item A edge with pattern $p(?i)$ is added from bound variable $i$ to sequence $s$ (resp. map $m$) if $s[p(i)]$ (resp. $m[p(i)]$) appears in the body of quantifier over $i$. The conditions for the edge are that associated to $m[p(i)]$ relative to the body of the quantified fact.
\end{itemize}
Note the conditions for edges may contain the bound variable $i$.

Initial instantiations at the current stage arise from \emph{free} appearances of terms of the form $a[i]$ for sequence $a$, and $m[k]$ or $\textsf{indom}(k, m)$ for map $m$. In such cases, $a$ (resp. $m$) already correspond to nodes in the graph, and $i$ (resp. $k$) are added as an initial instantiation for $a$ (resp. $m$). The conditions for those instantiations are that of $a[i]$ (resp. $m[k]$ or $\textsf{indom}(k, m)$) relative to the assumption or goal they appear in.

Instantiations are \emph{propagated} through the graph according to the following rules:

\begin{itemize}
\item For each instantiation $t$ on bound variable $i$, and edge from $i$ to sequence $a$ with weight $c$, add instantiation $t+c$ to $a$. Conditions for $t+c$ are formed by appending the conditions for $t$ with the conditions on the edge, with $i$ substituted by $t$.
\item Propagation along weighted edges can also proceed in reverse, from sequence $a$ to bound variable $i$. The added weight is $-c$. Conditions for $t-c$ are formed by appending the conditions for $t$ with conditions on the edge, with $i$ substituted by $t-c$.
\item For each instantiation $t$ on sequence $s$ (resp. map $m$), and edge from $i$ to $s$ (resp. map $m$) with pattern $p(?i)$, match $t$ against $p$. If there is a match with $?i:=t'$, add instantiation $t'$ for $i$. Conditions for $t'$ are formed by appending conditions for $t$ with conditions on the edge, with $i$ substituted by $s$.
\item For two sequence nodes (resp. map nodes) $n_1,n_2$ with the same name (as in Section~\ref{sec:generalized-atomic}), propagate instantiations from one to the other, with extra conditions stating equality between all corresponding indices.
\end{itemize}

As an example of the last propagation rule, given sequence nodes $a[i]$ and $a[j]$ (that is, $a$ is a nested sequence, so both $a[i]$ and $a[j]$ are sequences), propagate any instantiation $t$ from node $a[i]$ to node $a[j]$ with extra condition $i=j$.

We justify the rules for adding conditions as follows. In the first three propagation rules, the added conditions are the conditions for the term that created the edge relative to the body of the quantifier, \emph{with the bound variable instantiated to its value either in the source or target of the propagation}. In the last propagation rule, the intuition is that for any instantiation on $n_1$, the same instantiation on $n_2$ is necessary in situations where the indices are equal (so they are actually evaluated to the same sequence or map).

Whenever an instantiation is added according to the above rules, the list of conditions is combined with the quantifier-free facts in the current proof goal, and checked for inconsistency using the backend SMT solver. If the conditions are found to be inconsistent, the instantiation is not added (so the propagation is cut off here). This prevents some commonly-arising instantiation loops, as we show in Section~\ref{sec:examples}.

Each round of instantiation removes a single quantifier from each quantified fact. The quantified fact is removed from the proof state, and one fact is added back for each instantiation of the outermost bound variable. As there may be nested quantifiers, and new terms are produced as the result of instantiations, multiple rounds of instantiations may be required. The classification graph is preserved across the multiple rounds, and in each round, new instantiations may arise from either newly added nodes or newly added edges.

In the above procedure, instantiation loops may arise either in a single round, or due to the number of rounds being unbounded. We place cutoffs on the instantiation loops by limiting both the number of instantiation in each round, and the \emph{generation number} during instantiation. The generation number for terms is computed as follows:
\begin{itemize}
\item Terms in the original proof state have generation 0.
\item If a term first occurs in a quantifier-free fact that is the result of instantiation from terms of generation $i$, then it has generation $i+1$.
\end{itemize}

We keep track of generation number along propagations. A \emph{cutoff} $k$ is placed on the generation number. That is, terms with generation $k$ are no longer added to the graph (and hence cannot produce new instantiations). For our experiments we choose $k=2$.

From the above description, it will be noted that we do not add initial instantiations for conditions of the form $i\le c$ in the body of a quantified fact, nor $-1$ and $|a|$ for each sequence $a$, as are done in~\cite{DBLP:journals/jar/WangA23}. While these instantiations are necessary for completeness, we observe that they are rarely useful in proof goals that arise in OS verification, and can slow down (or even create loops) in the quantifier instantiation process. We show one example of this in Section~\ref{sec:examples}. 

\subsection{SMT Encoding}
\label{sec:smt-encoding}

The third step of the algorithm converts the quantifier-free formula that results from quantifier instantiation into a form suitable for SMT solvers.

For structures and enumerations, a common choice for hammer-like tools in proof assistants is to declare a new type for the structure or enumeration, and encode field access as functions mapping the new type into the field type. This has the advantage of being general, in particular applicable to inductively defined datatypes. However, they also create problems in that various axioms about the datatype need to be added as lemmas to the SMT solver, including definition of equality on the datatype, rewriting rules for field access, etc. The axioms themselves have quantifiers which can complicate SMT reasoning.

We employ an alternative approach in \textsf{OSVAuto}, encoding each name of a generalized atomic term as a variable or uninterpreted function in the SMT formula. This works because we do not allow recursion in  datatype definitions. Moreover, due to Theorem~\ref{thm:normalization}, all generalized atomic terms at this stage have primitive types.

More precisely, for each generalized atomic term $t$, decomposed it into the name-indices pair $(n, [i_1, \dots, i_k])$ according to Section~\ref{sec:generalized-atomic}. Encode $n$ as an SMT variable or uninterpreted function $f_{n}$ with type $\textsf{type}(i_1)\Rightarrow \dots \Rightarrow \textsf{type}(i_k)\Rightarrow \textsf{type}(t)$. The term $t$ is then encoded as the SMT term $t(i_1',\dots,i_k')$, where $i_j'$ is the recursively computed encoding for $i_j$ for $1\le j\le k$. The logical and arithmetic operators are converted to corresponding SMT operations in the standard way.

\begin{example}
\label{ex:encoding}
Continuing from Example~\ref{ex:normalize}. Designate SMT variables $n_1,\dots,n_{12}$ for names \textsf{s.id}, \textsf{t.id}, \textsf{s.pts.length}, \textsf{t.pts.length}, \textsf{s.pts.index.x}, \textsf{s.pts.index.y}, \textsf{t.pts.index.x}, \textsf{t.pts.index.y}, 
\textsf{s.pt.x}, \textsf{s.pt.y}, \textsf{t.pt.x} and \textsf{t.pt.y}. Here $n_5$ through $n_8$ are uninterpreted functions on integers. The equality $s=t$ is encoded as:
\begin{align*}
&n_1 = n_2\,\wedge \\
&(n_1 = 0 \longrightarrow n_3 = n_4 \wedge \forall i.\ 0\le i<n_3 \longrightarrow n_5(i) = n_7(i) \wedge n_6(i) = n_8(i)) \,\wedge \\
&(n_1 = 1 \longrightarrow n_9 = n_{11} \wedge n_{10} = n_{12})
\end{align*}
(although in practice, only versions of the above with $i$ instantiated to concrete values will be encoded).
\end{example}

\subsection{Examples}
\label{sec:examples}

In this section, we give some examples showing the algorithm in action. The examples justify our use of conditions during propagation, omission of instantiations from inequality conditions, and propagation between nodes with the same name.

\subsubsection{Variant of the \textsf{append} example}

We first study a variant of the running example in \cite{DBLP:journals/jar/WangA23}, as reproduced in Section~\ref{sec:sequence-preliminary}. The original example makes use of the fact that equations $c=\textsf{append}(a, b)$ and $d=\textsf{append}(b,a)$ can be used for rewriting. We consider a variant where $a, b, c, d$ are all values of some map, as follows: fix $g$ of type \textsf{Map<int, Seq<int32u>}\textsf{>}. Assume $g[2] = \textsf{append}(g[0], g[1])$ and $g[3] = \textsf{append}(g[1], g[0])$, then show:
\[ (\forall i.\, 0\le i<|p[2]| \rightarrow P(p[2][i])) \Longrightarrow (\forall k.\, 0\le k<|p[3]| \rightarrow P(p[3][k])) \]
Unlike before, since $g[2]$ and $g[3]$ have indices, they cannot be used as rewrite rules. The assumption $g[2] = \textsf{append}(g[0], g[1])$ expands into:
\begin{align*}
\forall i.\ & (0\le i < |g[0]| \longrightarrow g[2][i] = g[0][i]) \ \wedge \\
& (|g[0]|\le i < |g[0]| + |g[1]| \longrightarrow g[2][i] = g[1][i - |g[0]|])
\end{align*}
Likewise, the assumption $g[3] = \textsf{append}(g[1], g[2])$ expands into:
\begin{align*}
\forall j.\ & (0\le j < |g[1]| \longrightarrow g[3][j] = g[1][j]) \ \wedge \\
& (|g[1]|\le j < |g[1]| + |g[0]| \longrightarrow g[3][j] = g[0][j - |g[1]|])
\end{align*}

We then get nodes $i, j, g[0], g[1], g[2], g[3]$ in the graph, with the following edges: $i\to g[0]$ and $i\to g[2]$ with weight 0; $i\to g[1]$ with weight $-|g[0]|$; $j\to g[1]$, $j\to g[3]$ with weight 0; $j\to g[0]$ with weight $-|g[1]|$. This graph has a cycle
\[ i\to g[0] \to j \to g[1] \to i \]
with total weight $|g[1]| + |g[0]|$ (in~\cite{DBLP:journals/jar/WangA23} the assumptions would be split, this does not affect the determination of cycles).

In~\cite{DBLP:journals/jar/WangA23}, this non-zero cycle would prevent instantiation from converging. However, with the help of conditions, we will see the cycle is spurious. Similar to before, the conclusion reduces to $P(g[3][k])$ with condition $0\le k<|g[3]|$. This gives instantiation $k$ for node $g[3]$ under condition $0\le k < |g[3]|$. This instantiation is first propagated to node $j$. When further propagated along edges $g[1]$, $i$, and $g[0]$, eventually a condition $0\le k+|g[0]|<|g[0]|$ arises, which implies $k<0$, contradicting the condition $0\le k$ earlier. Hence, the last instantiation is not added and propagation stops here. Thus the instantiation converges and the resulting quantifier-free statement is provable.

\subsubsection{Reasoning about uniqueness}

It is common in program verification to encounter \emph{uniqueness} conditions -- that is any two elements of a sequence is different. They can be stated as follows:
\[
\textsf{unique}(a) = \forall i\ j.\ 0\le i<|a| \wedge 0\le j<|a| \wedge i\neq j \longrightarrow a[i]\neq a[j]
\]
Such uniqueness conditions are not covered by the algorithm in~\cite{DBLP:journals/jar/WangA23}: the condition $i\neq j$ will produce edges with weight 1 and $-1$ between bound variables $i$ and $j$, which give rise to nonzero cycles (see also the analysis in Section~\ref{sec:discussion}).

In our case, since we omitted instantiation on inequality conditions, we only add edges $i\to a$ and $j\to a$ with weight 0, which themselves do not give rise to nonzero cycles. We illustrate with the following example.

Given $a$ with type $\textsf{Seq<T>}$, integer $k$ satisfying $0\le k<|a|$, and let $b=\textsf{remove}(k, a)$ (removing the $k$'th element from $a$). We wish to show that $a$ has unique elements, then $b$ does not contain the removed element $a[k]$:
\[
\textsf{unique}(a) \Longrightarrow \forall m.\ 0\le m<|b| \longrightarrow b[m] \neq a[k]
\]
First, substitute $\textsf{remove}(k, a)$ for $b$ in the goal, then expand the definition of $\textsf{remove}$. This results in:
\begin{align*}
0\le m<\textsf{max}(0, |a|-1) \Longrightarrow
&(0\le m < k \longrightarrow a[m] \neq a[k])\ \wedge \\
&(k\le m < |a| - 1 \longrightarrow a[m+1] \neq a[k])
\end{align*}

In the first round, nodes for bound variable $i$ and sequence $a$ are created, with edge $i\to a$ with weight 0. The initial instantiations for $a$ are $m, m+1, k$. This propagates to instantiations $m, m+1, k$ for $i$. Thus, $\textsf{unique}(a)$ give rise to three quantified facts over $j$.


In the second round, nodes for bound variables $j_1, j_2, j_3$ are created, with edges $j_1\to a$, $j_2\to a$, $j_3\to a$ with weight 0. The following instantiations are propagated: $k$ for $j_1$ and $j_2$, and $m, m+1$ for $j_3$. This results in four quantifier-free facts, which are sufficient for proving the goal.

\subsubsection{Propagation across nested sequences}

In this example, we demonstrate the importance of propagation between sequence or map nodes with the same name but different indices. The following predicate states that all rows have unique elements for a nested sequence:
\[
\textsf{rows\_unique}(a) = \forall k.\ 0\le k<|a| \longrightarrow \textsf{unique}(a[k])
\]
Now given a nested sequence $a$, and $b$ is formed by removing the first element of $a[x]$ ($0\le x\le |a|$), which expands into the following conditions:
\[
|b| = |a| \wedge b[x] = \textsf{remove}(0, a[x]) \wedge \forall y.\ 0\le y<|a| \wedge y\neq x \longrightarrow b[y] = a[y]
\]
We wish to show that under these conditions,
$\textsf{row\_unique(a)} \longrightarrow \textsf{row\_unique(b)}$.

The full computation is rather long and we will explain only the key parts. From the goal, we introduce variables $k,i,j$, assumption $i\neq j$, and goal $b[k][i]\neq b[k][j]$. The condition $b[x]=\textsf{remove}(0, a[x])$ expands into:
\[
\forall m.\ 0\le m<|a|-1 \longrightarrow b[x][m] = a[x][m+1]
\]
In the first round of instantiation, node $b[k]$ is created with initial instantiations $i$ and $j$. Propagating these instantiations across sequence/map nodes with the same name, we obtain instantiations $i$ and $j$ for $b[x]$ (under condition $x=k$). These in turn propagate to instantiations $i, j, i-1, j-1$ for the bound variable $m$, which are necessary for proving the goal. The instantiation eventually converges, yielding a provable quantifier-free statement.

\section{Case study}
\label{sec:case-study}

We implemented the functional specification language and algorithm described in Section~\ref{sec:language} and~\ref{sec:algorithm} as the tool \textsf{OSVAuto} using Python. We use \textsf{z3}~\cite{DBLP:conf/tacas/MouraB08} as the SMT backend accessed through its Python API. In addition to some standard test cases (including the examples explained in Section~\ref{sec:examples}), we mainly evaluate our tool using the existing verification of the $\mu$C/OS-II kernel in Coq\footnote{Source files and examples can be found at \url{https://anonymous.4open.science/r/OSVAuto-8F62}.}.

The basic structure of the verification is already explained in Section~\ref{sec:ucos-preliminary}. We emphasize again that the current work deals only with verification of functional properties, with combination with (semi-)automated reasoning about concurrent separation logic left to future work. There are several aspects in the design of $\mu$C/OS-II that are quite intricate, where verification of functional properties requires rather lengthy reasoning:

\begin{itemize}
\item Readiness of tasks at each priority is stored both in the TCB itself, as well as in a priority table for more efficient retrieval. They need to be updated consistently whenever task status changes.

\item Data about events is stored in both event control blocks and task control blocks. They must be updated in a consistent whenever events are sent or received.

\item The list of available messages in a message queue is stored in a circular buffer. The refinement relation between the low-level data structure (a circular array) and the high-level data (the actual list of messages represented) is quite complex.

\end{itemize}

Nested sequences and maps are used in two places in our model: for each event, the priority table storing list of tasks waiting for the event is a sequence, and for message queue events, the list of messages in the queue is a sequence.

The following table summarizes API functions that are modeled in this work. For task-related APIs, we verify preservation of invariants on TCB and priority table, and refinement relations between low-level and high-level TCBs. For event-related APIs, we verify consistency of event information in the data structures, and refinement relation between low-leven and high-level event control blocks. For message queues (the three \textsf{OSQ} functions), the correctness of the circular buffer is verified. We emphasize that all proof obligations are verified \emph{automatically}, with the user only having to state goals on invariant and refinement preservation.

\begin{table}
\centering
\begin{tabular}{|c|c|c|c|}
\hline
\hspace{0.2cm}\textbf{Function}\hspace{0.2cm} & 
\hspace{0.2cm}\textbf{\#Lines Spec}\hspace{0.2cm} &
\hspace{0.2cm}\textbf{\#Goals}\hspace{0.2cm} &
\hspace{0.2cm}\textbf{Total Time (s)}\hspace{0.2cm} \\
\hline
OSTaskSuspend & 18 & 3 & 2.3 \\
\hline
OSTaskResume & 28 & 3 & 2.5 \\
\hline
OSSemPend & 38 & 8 & 11.0 \\
\hline
OSSemPost & 38 & 8 & 50.7 \\
\hline
OSMboxPend & 50 & 8 & 11.9 \\
\hline
OSMboxPost & 40 & 8 & 51.8 \\
\hline
OSQAccept & 29 & 4 & 12.4 \\
\hline
OSQPend & 52 & 8 & 19.6 \\
\hline
OSQPost & 46 & 8 & 56.5 \\
\hline
 \hspace{0.2cm} Auxiliary+Invariants \hspace{0.2cm} & 125+437 & & \\
\hline
Total & 901 & 58 & 218.7 \\
\hline
\end{tabular}
\caption{Summary of results from the $\mu$C/OS-II case study. For each API function that are modeled, we list number of lines of specification of the function, number of proof obligations checked, and total running time. We also list number of lines of auxiliary definitions and invariants.}
\label{tab:case-study}
\end{table}

\section{Discussion}
\label{sec:discussion}

In~\cite{DBLP:journals/jar/WangA23}, a proof of completeness is given for the quantifier instantiation method proposed there. The main idea of the proof is to construct a counterexample for the original goal by extending the model returned by the SMT solver, such that for each sequence, the same value is used for each entry in any interval cut out by the instantiated indexes. We give two examples showing why it is difficult to extend the proof to cover more general situations that arise in OS verification.
\begin{enumerate}
\item It is often necessary to express that a sequence contains distinct values, the method for building counterexamples above is inapplicable in this case (the same applies to the strictly-sorted condition).
\item Sometimes each entry of a sequence must satisfy additional properties related to its index. For example, it is common to require that for the mapping \textsf{tcbMap} from priority to TCBs, the condition $\forall i.\ \textsf{tcbMap}[i].\textsf{prio} = i$ is needed. Again the counterexample does not work here as each value of the map must be different under this requirement.
\end{enumerate}
We leave completeness proofs (which would require modification of the quantifier instantiation algorithm and/or restrictions to the setting in which it applies) to future work.

\section{Related work}
\label{sec:related-work}

Most related to our work are existing research on decision procedures for arrays and sequences. The array property fragment is first proposed in Bradley et al.~\cite{DBLP:conf/vmcai/BradleyMS06}. The work of Ge and de Moura~\cite{DBLP:conf/cav/GeM09} generalized this fragment to include index offsets. The algorithm of our work builds upon that of~\cite{DBLP:journals/jar/WangA23}, which also provides a nice review of existing work. Other recent work on sequence theory include~\cite{DBLP:conf/cav/JezLMR23}, which focused regular constraints on sequences modeled using automata. The resulting tool \textsf{SeCo} is applied to verify protocols such as Lamport's Bakery Algorithm and Dijkstra's Self-Stabilizing Protocol. The work~\cite{DBLP:journals/jar/ShengNRZDGPQBT23} proposed a number of reasoning rules about sequences, and applied them to proof obligations from verifying smart contracts using Diem. However, none of these work address the case where elements in the sequence can themselves be sequences (that is, nested sequences), nor combination with structure and enumeration definitions. We focus on extension of automation techniques to this case, addressing a need that arises in more complicated application scenarios.


The difficulty with verification of operating systems by interactive theorem proving have led researchers to consider more automatic methods in the recent years. In particular, there have been attempts at fully-automatic (push-button) verification~\cite{DBLP:conf/sosp/NelsonSZJBTW17,DBLP:conf/sosp/NelsonBGBTW19}. There are also semi-automatic verification tools for C~\cite{DBLP:conf/pldi/SammlerLKMD021} and Rust~\cite{DBLP:journals/pacmpl/LattuadaHCBSZHPH23}. The tool VST-A~\cite{DBLP:journals/pacmpl/ZhouQWAC24} improves automation in the Verified Software Toolchain~\cite{DBLP:conf/esop/Appel11} by allowing users to perform verification using annotations. Research on these tools focused mostly on the program logic involved. Our work can be seen as complementary, focusing final stage of using SMT to solve the proof obligations.


\section{Conclusion}
\label{sec:conclusion}

In this paper, we propose proof automation techniques for verifying functional specifications that commonly arise in OS verification. We mainly address issues of quantifier instantiation when nested sequences and maps are present. The algorithm is implemented in a tool \textsf{OSVAuto}, and evaluated on the existing verification of $\mu$C/OS-II kernel in Coq. The results show that the algorithm perform well on the lengthy proof obligations that arise, with the potential to significantly reduce proof effort when combined with other tools for reasoning about separation logic.

\bibliographystyle{plain}

\begin{thebibliography}{10}

\bibitem{DBLP:conf/esop/Appel11}
Andrew~W. Appel.
\newblock Verified software toolchain - (invited talk).
\newblock In Gilles Barthe, editor, {\em Programming Languages and Systems -
  20th European Symposium on Programming, {ESOP} 2011, Held as Part of the
  Joint European Conferences on Theory and Practice of Software, {ETAPS} 2011,
  Saarbr{\"{u}}cken, Germany, March 26-April 3, 2011. Proceedings}, volume 6602
  of {\em Lecture Notes in Computer Science}, pages 1--17. Springer, 2011.

\bibitem{DBLP:conf/vmcai/BradleyMS06}
Aaron~R. Bradley, Zohar Manna, and Henny~B. Sipma.
\newblock What's decidable about arrays?
\newblock In E.~Allen Emerson and Kedar~S. Namjoshi, editors, {\em
  Verification, Model Checking, and Abstract Interpretation, 7th International
  Conference, {VMCAI} 2006, Charleston, SC, USA, January 8-10, 2006,
  Proceedings}, volume 3855 of {\em Lecture Notes in Computer Science}, pages
  427--442. Springer, 2006.

\bibitem{DBLP:conf/tacas/MouraB08}
Leonardo~Mendon{\c{c}}a de~Moura and Nikolaj~S. Bj{\o}rner.
\newblock {Z3:} an efficient {SMT} solver.
\newblock In C.~R. Ramakrishnan and Jakob Rehof, editors, {\em Tools and
  Algorithms for the Construction and Analysis of Systems, 14th International
  Conference, {TACAS} 2008, Held as Part of the Joint European Conferences on
  Theory and Practice of Software, {ETAPS} 2008, Budapest, Hungary, March
  29-April 6, 2008. Proceedings}, volume 4963 of {\em Lecture Notes in Computer
  Science}, pages 337--340. Springer, 2008.

\bibitem{DBLP:conf/cav/EkiciMTKKRB17}
Burak Ekici, Alain Mebsout, Cesare Tinelli, Chantal Keller, Guy Katz, Andrew
  Reynolds, and Clark~W. Barrett.
\newblock Smtcoq: {A} plug-in for integrating {SMT} solvers into coq.
\newblock In Rupak Majumdar and Viktor Kuncak, editors, {\em Computer Aided
  Verification - 29th International Conference, {CAV} 2017, Heidelberg,
  Germany, July 24-28, 2017, Proceedings, Part {II}}, volume 10427 of {\em
  Lecture Notes in Computer Science}, pages 126--133. Springer, 2017.

\bibitem{DBLP:conf/cav/GeM09}
Yeting Ge and Leonardo~Mendon{\c{c}}a de~Moura.
\newblock Complete instantiation for quantified formulas in satisfiabiliby
  modulo theories.
\newblock In Ahmed Bouajjani and Oded Maler, editors, {\em Computer Aided
  Verification, 21st International Conference, {CAV} 2009, Grenoble, France,
  June 26 - July 2, 2009. Proceedings}, volume 5643 of {\em Lecture Notes in
  Computer Science}, pages 306--320. Springer, 2009.

\bibitem{DBLP:conf/osdi/GuSCWKSC16}
Ronghui Gu, Zhong Shao, Hao Chen, Xiongnan~(Newman) Wu, Jieung Kim, Vilhelm
  Sj{\"{o}}berg, and David Costanzo.
\newblock Certi{KOS}: An extensible architecture for building certified
  concurrent os kernels.
\newblock In Kimberly Keeton and Timothy Roscoe, editors, {\em 12th {USENIX}
  Symposium on Operating Systems Design and Implementation, {OSDI} 2016,
  Savannah, GA, USA, November 2-4, 2016}, pages 653--669. {USENIX} Association,
  2016.

\bibitem{DBLP:conf/cav/JezLMR23}
Artur Jez, Anthony~W. Lin, Oliver Markgraf, and Philipp R{\"{u}}mmer.
\newblock Decision procedures for sequence theories.
\newblock In Constantin Enea and Akash Lal, editors, {\em Computer Aided
  Verification - 35th International Conference, {CAV} 2023, Paris, France, July
  17-22, 2023, Proceedings, Part {II}}, volume 13965 of {\em Lecture Notes in
  Computer Science}, pages 18--40. Springer, 2023.

\bibitem{DBLP:journals/tocs/KleinAEMSKH14}
Gerwin Klein, June Andronick, Kevin Elphinstone, Toby~C. Murray, Thomas Sewell,
  Rafal Kolanski, and Gernot Heiser.
\newblock Comprehensive formal verification of an {OS} microkernel.
\newblock {\em {ACM} Trans. Comput. Syst.}, 32(1):2:1--2:70, 2014.

\bibitem{DBLP:journals/pacmpl/LattuadaHCBSZHPH23}
Andrea Lattuada, Travis Hance, Chanhee Cho, Matthias Brun, Isitha Subasinghe,
  Yi~Zhou, Jon Howell, Bryan Parno, and Chris Hawblitzel.
\newblock Verus: Verifying rust programs using linear ghost types.
\newblock {\em Proc. {ACM} Program. Lang.}, 7({OOPSLA1}):286--315, 2023.

\bibitem{DBLP:conf/sosp/NelsonBGBTW19}
Luke Nelson, James Bornholt, Ronghui Gu, Andrew Baumann, Emina Torlak, and
  Xi~Wang.
\newblock Scaling symbolic evaluation for automated verification of systems
  code with serval.
\newblock In Tim Brecht and Carey Williamson, editors, {\em Proceedings of the
  27th {ACM} Symposium on Operating Systems Principles, {SOSP} 2019,
  Huntsville, ON, Canada, October 27-30, 2019}, pages 225--242. {ACM}, 2019.

\bibitem{DBLP:conf/sosp/NelsonSZJBTW17}
Luke Nelson, Helgi Sigurbjarnarson, Kaiyuan Zhang, Dylan Johnson, James
  Bornholt, Emina Torlak, and Xi~Wang.
\newblock Hyperkernel: Push-button verification of an {OS} kernel.
\newblock In {\em Proceedings of the 26th Symposium on Operating Systems
  Principles, Shanghai, China, October 28-31, 2017}, pages 252--269. {ACM},
  2017.

\bibitem{DBLP:conf/pldi/SammlerLKMD021}
Michael Sammler, Rodolphe Lepigre, Robbert Krebbers, Kayvan Memarian, Derek
  Dreyer, and Deepak Garg.
\newblock Refinedc: automating the foundational verification of {C} code with
  refined ownership types.
\newblock In Stephen~N. Freund and Eran Yahav, editors, {\em {PLDI} '21: 42nd
  {ACM} {SIGPLAN} International Conference on Programming Language Design and
  Implementation, Virtual Event, Canada, June 20-25, 2021}, pages 158--174.
  {ACM}, 2021.

\bibitem{DBLP:journals/jar/ShengNRZDGPQBT23}
Ying Sheng, Andres N{\"{o}}tzli, Andrew Reynolds, Yoni Zohar, David~L. Dill,
  Wolfgang Grieskamp, Junkil Park, Shaz Qadeer, Clark~W. Barrett, and Cesare
  Tinelli.
\newblock Reasoning about vectors: Satisfiability modulo a theory of sequences.
\newblock {\em J. Autom. Reason.}, 67(3):32, 2023.

\bibitem{DBLP:journals/jar/WangA23}
Qinshi Wang and Andrew~W. Appel.
\newblock A solver for arrays with concatenation.
\newblock {\em J. Autom. Reason.}, 67(1):4, 2023.

\bibitem{DBLP:conf/cav/XuFFZZL16}
Fengwei Xu, Ming Fu, Xinyu Feng, Xiaoran Zhang, Hui Zhang, and Zhaohui Li.
\newblock A practical verification framework for preemptive {OS} kernels.
\newblock In Swarat Chaudhuri and Azadeh Farzan, editors, {\em Computer Aided
  Verification - 28th International Conference, {CAV} 2016, Toronto, ON,
  Canada, July 17-23, 2016, Proceedings, Part {II}}, volume 9780 of {\em
  Lecture Notes in Computer Science}, pages 59--79. Springer, 2016.

\bibitem{DBLP:journals/pacmpl/ZhouQWAC24}
Litao Zhou, Jianxing Qin, Qinshi Wang, Andrew~W. Appel, and Qinxiang Cao.
\newblock {VST-A:} {A} foundationally sound annotation verifier.
\newblock {\em Proc. {ACM} Program. Lang.}, 8({POPL}):2069--2098, 2024.

\end{thebibliography}

\end{document}